\documentclass{rr-rspublic}
\usepackage{epsfig,graphicx,natbib,txfonts,url,twoopt}
\usepackage[breaklinks=true]{hyperref}  %RR to avoid \cite line fills

\bibpunct{(}{)}{;}{a}{}{,}    %% natbib format choice like A&A and ApJ
\def\cite#1{\citealp{#1}}     %% restore old astroncite \cite command
\newcommandtwoopt{\citeads}[3][][]{\href{http://adsabs.harvard.edu/abs/#3}%
                                        {\citealp[#1][#2]{#3}}}
\newcommandtwoopt{\citepads}[3][][]{\href{http://adsabs.harvard.edu/abs/#3}%
                                         {\citep[#1][#2]{#3}}}
\newcommandtwoopt{\citetads}[3][][]{\href{http://adsabs.harvard.edu/abs/#3}%
                                         {\citet[#1][#2]{#3}}}
\newcommandtwoopt{\citeyearads}%
  [3][][]{\href{http://adsabs.harvard.edu/abs/#3}%
  {\citeyear[#1][#2]{#3}}}
\newcount\longrefs
\def\aap{\ifnum\longrefs=1 {Astron.\ Astrophys.}\else 
                           {A\hbox{\rm \&}A}\fi}
\def\aapr{\ifnum\longrefs=1 {Astron.\ Astrophys.\ Rev.}\else 
                            {A\hbox{\rm \&}AR}\fi}
\def\aaps{\ifnum\longrefs=1 {Astron.\ Astrophys.\ Suppl.}\else 
                            {A\hbox{\rm \&}A Suppl.}\fi}
\def\aj{\ifnum\longrefs=1 {Astron.\ J.}\else 
                          {AJ}\fi} 
\def\ao{\ifnum\longrefs=1 {Applied Optics}\else 
                           {Appl.\ Opt.}\fi} 
\def\aspcs{\ifnum\longrefs=1 {Astron.\ Soc.\ Pacific Conf. Series}\else 
                           {ASP Conf.\ Ser.}\fi} 
\def\apj{\ifnum\longrefs=1 {Astrophys.\ J.}\else 
                           {ApJ}\fi} 
\def\apjl{\ifnum\longrefs=1 {Astrophys.\ J. Lett.}\else 
                            {ApJ}\fi} 
\def\aplett{\ifnum\longrefs=1 {Astrophys.\ J. Lett.}\else 
                            {ApJ}\fi} 
\def\apjs{\ifnum\longrefs=1 {Astrophys.\ J. Suppl.}\else 
                            {ApJS}\fi}
\def\apss{\ifnum\longrefs=1 {Astrophys.\ and Space Science}\else 
                            {Astrophys.\ Space Sci.}\fi}
\def\araa{\ifnum\longrefs=1 {Ann.\ Rev.\ Astron.\ Astrophys.}\else 
                            {ARA\hbox{\rm \&}A}\fi}
\def\azh{\ifnum\longrefs=1 {Astronomicheskii Zhurnal}\else 
                            {Astron.\ Zhur.}\fi}
\def\baas{\ifnum\longrefs=1 {Bull.\ Am.\ Astron.\ Soc.}\else 
                            {BAAS}\fi}
\def\bain{\ifnum\longrefs=1 {Bull.\ Astronom.\ Institutes Netherlands}\else
                            {Bull.\ Astr.\ Inst.\ Neth.}\fi}
\def\gca{\ifnum\longrefs=1 {Geochim.\ Cosmochim.\ Acta}\else 
                           {Geochim.\ Cosmochim.\ Acta}\fi}
\def\grl{\ifnum\longrefs=1 {Geophys.\ Res.\ Lett.}\else 
                           {Geoph.\ Res.\ Lett.}\fi}
\def\iaucirc{\ifnum\longrefs=1 {IAU Circulars}\else 
                          {IAU Circ.}\fi}
\def\ip{\ifnum\longrefs=1 {in press}\else 
                          {in press}\fi}
\def\jgr{\ifnum\longrefs=1 {J.\ Geophys.\ Res.}\else 
                           {J.\ Geophys.\ Res.}\fi}  
\def\jrasc{\ifnum\longrefs=1 {J.\ Royal Astron.\ Soc.\ Canada}\else 
                           {JRAS Can.}\fi}  
\def\memsai{\ifnum\longrefs=1 {Mem.~Soc.~Astron.~Italiana}\else
                              {MemSAI}\fi}
\def\mnras{\ifnum\longrefs=1 {Mon.\ Not.\ Roy.\ Astron.\ Soc.}\else 
                             {MNRAS}\fi} 
\def\nat{\ifnum\longrefs=1 {Nature}\else 
                           {Nat}\fi}
\def\pasj{\ifnum\longrefs=1 {Pub.\ Astron.\ Soc.\ Japan}\else 
                            {PASJ}\fi} 
\def\pasp{\ifnum\longrefs=1 {Pub.\ Astron.\ Soc.\ Pacific}\else 
                            {PASP}\fi} 
\def\physscr{\ifnum\longrefs=1 {Physica Scripta}\else 
                            {Phys.\ Scrip.}\fi} 
\def\planss{\ifnum\longrefs=1 {Planetary \& Space Science}\else 
                            {Plan. \& Space Sci.}\fi} 
\def\procspie{\ifnum\longrefs=1 {Proc.\ SPIE}\else 
                            {Proc.\ SPIE}\fi} 
\def\qjras{\ifnum\longrefs=1 {Quarterly J.\ Royal Astron.\ Soc.}\else 
                            {QJRAS}\fi} 
\def\sa{\ifnum\longrefs=1 {Soviet Astron..}\else 
                               {Sov.\ Astron.}\fi}
\def\skytel{\ifnum\longrefs=1 {Sky \& Telescope}\else 
                            {Sky \& Tel.}\fi} 
\def\solphys{\ifnum\longrefs=1 {Solar Phys.}\else 
                               {Sol.\ Phys.}\fi}
\def\ssr{\ifnum\longrefs=1 {Space Science Rev.}\else 
                               {Space\ Sci.\ Rev.}\fi}
\def\zap{\ifnum\longrefs=1 {Zeitschr.\ f.\ Astrophysik}\else
                               {Z.\ Astrophys.}\fi}

\def\nl{,\ } %%\def\nl{\newline}  %% redefine as \newline for mail addresses

\def\ITA{Institute of Theoretical Astrophysics\nl
         University of Oslo\nl
         P.O. Box 1029, Blindern\nl N--0315 Oslo\nl Norway}

\def\LMSAL{Lockheed-Martin Solar and Astrophysics Laboratory\nl
           3251 Hanover Street\nl Palo Alto, CA~94304\nl USA}

\def\SIU{Sterrekundig Instituut\nl Utrecht University\nl Postbus 80\,000\nl
         NL--3508~TA~Utrecht\nl The~Netherlands}

\long\def\startignore #1\stopignore{}   %% use \startignore....\stopignore

\def\rmit#1{{\it #1}}              %% italics (RR style, Kluwer)
\def\etc{\rmit{etc.}}           
\def\ie{\rmit{i.e.,}}              %% , required (Webster 1681)
\def\eg{\rmit{e.g.,}}              %% , required (Webster 1681)
\def\cf{cf.}                       %% no Latin, always Roman (Webster 1686)

\def\specchar#1{\uppercase{#1}}    %% to be redefined for A&A, small caps
\def\CaII{\mbox{Ca\,\specchar{ii}}}

\def\FeI{\mbox{Fe\,\specchar{i}}}

\def\HI{\mbox{H\,\specchar{i}}} 
 
\def\Hmin{\hbox{\rmH$^{^{_-}}\!$}}      %% H^min, very elegant
\def\HeII{\mbox{He\,\specchar{ii}}}

\def\NaI{\mbox{Na\,\specchar{i}}}

\def\Halpha{\mbox{H\hspace{0.1ex}$\alpha$}} %% \Halpha\ for space behind it

\def\Lyalpha{\mbox{Ly$\hspace{0.2ex}\alpha$}}

\def\Done{\mbox{D$_1$}}

\def\CaIIH{\mbox{Ca\,\specchar{ii}\,\,H}}

\def\HK{\mbox{H\,\&\,K}}
\def\HtwoV{\mbox{H$_{2V}$}}

 \def\rmH{{\rm H}}

\def\arcsec{\hbox{$^{\prime\prime}$}}

\def\is{\!=\!}                             %% tighter spacing
\def\={\hbox{$\!=\!$}}                     %% no space around =

\def\figspath{.}
\def\topic#1{\par \vspace{1ex} \noindent {\em #1.\,}}
\def\rrsubsection#1{\subsection{#1}\mbox{}\vspace{-4.5ex}}

\begin{document}

\title[Photosphere and Chromosphere]
      {The Quiet-Sun Photosphere and Chromosphere}
\author[R.J. Rutten]{Robert J. Rutten}
\affiliation{\SIU \\ \ITA\\ \LMSAL}

%RR bad instruction, should be single postal address

\label{firstpage}
\maketitle

\begin{abstract}{Sun, photosphere, chromosphere}
  The overall structure and the fine structure of the solar
  photosphere outside active regions are largely understood, except
  possibly important roles of a turbulent near-surface dynamo at its
  bottom, internal gravity waves at its top, and small-scale
  vorticity.  Classical 1D static radiation-escape modelling has been
  replaced by 3D time-dependent MHD simulations that come closer to
  reality.

  The solar chromosphere, in contrast, remains ill-understood although
  its pivotal role in coronal mass and energy loading makes it a
  principal research area.  Its fine structure defines its overall
  structure, so that hard-to-observe and hard-to-model small-scale
  dynamical processes are the key to understanding.  However, both
  chromospheric observation and chromospheric simulation presently mature
  towards the required sophistication.  The open-field features seem
  of greater interest than the easier-to-see closed-field features.
\end{abstract}

%%%%%%%%%%%%%%%%%%%%%%%%%%%%%%%%%%%%%%%%%%%%%%%%%%%%%%%%%%%%%%%%%%%%%%%%%%%%
\section{Introduction}
%%%%%%%%%%%%%%%%%%%%%%%%%%%%%%%%%%%%%%%%%%%%%%%%%%%%%%%%%%%%%%%%%%%%%%%%%%%%

Solar photosphere--chromosphere--corona coupling is presently a
premier research topic in trying to understand our star's regulation
of our environment.
%RR next to dynamo, AR appearance and disappeance, NLFFF, cycle
Large strides forward are made thanks to three methodological
advances: (1) real-time and post-detection wavefront correction
enabling 0.1\arcsec\ resolution from meter-class optical telescopes,
% with the SST\footnote{Swedish 1-m Solar Telescope:
%   \url{http://www.solarphysics.kva.se}.}  setting the example; 
(2)
continuous multi-wavelength high-cadence monitoring from space, 
% where
% the SDO\footnote{Solar Dynamics Observatory:
%   \url{http://sdo.gsfc.nasa.gov}.}  has begun unprecedented data
% collection; 
(3) increasing realism of numerical simulations of
solar-atmosphere fine structure.
% \eg\ 2011A&A...531A.154G  Bifrost

This brief overview summarises the status, issues, and prospects in
studying the lower solar atmosphere away from active
regions.
More detailed recent reviews of relevance are those of solar magnetism
by \citetads{2006RPPh...69..563S}%  % Solanki++ review
\footnote{On-screen readers of the ArXiv pdf preprint may click
  on the year in a citation
  to open the corresponding ADS abstract page in a browser.},
chromosphere observations by
\citetads{2006ASPC..354..259J} % Judge SPO Steinfest
and me (\citeads{2007ASPC..368...27R}), % Rutten Coimbra review
chromosphere modelling by
\citetads{2007ASPC..368...49C}, % Carlsson Coimbra
stellar chromospheres by
\citetads{2008LRSP....5....2H}, % Hall stellar chromospheres
solar convection simulations 
%%(including abundance ramifications) 
by \citetads{2009LRSP....6....2N}, % Nordlund+Stein+Asplund convection
small-scale photospheric magnetism by
\citetads{2009SSRv..144..275D}, % de Wijn++ small-scale fields
magnetic photosphere-chromosphere coupling by
\citetads{2010mcia.conf..166S}, % Steiner Evershed
and of supergranulation by
\citetads{2010LRSP....7....2R}. % Rieutord+Rinconn supergranulation

{\em Quiet Sun\/} denotes those areas of the solar atmosphere where magnetic
activity is not obvious on the solar surface in wide-band optical
continuum images.

%RR put figs on 2nd page for pdf printout
%===========================================================================
\begin{figure}
  \centerline{\includegraphics[width=\textwidth]{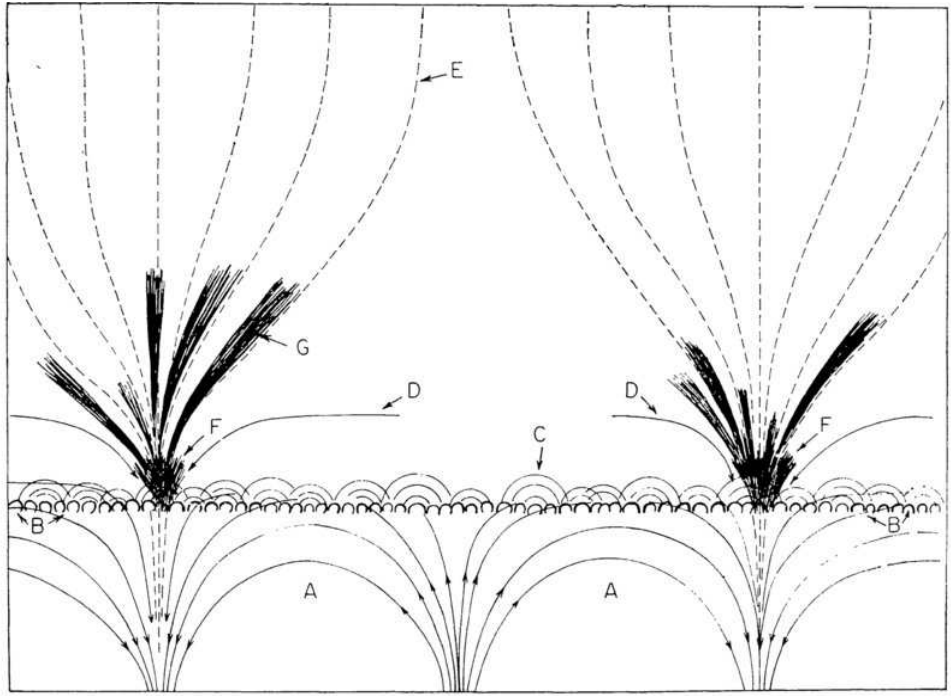}}
  \caption[]{\label{fig:Noyes}%
    Sketch of the granulation-supergranulation-spicule complex in
    cross-section.  A: flow lines of a supergranulation cell. B:
    photospheric granules. C: wave motions.  D: large-scale
    chromospheric flow field seen in \Halpha. E: [magnetic] lines of
    force, pictured as uniform in the corona but concentrated at the
    boundaries of the supergranules in the photosphere and
    chromosphere.  F: base of a spicule `bush' or `rosette', visible as
    a region of enhanced emission in the \Halpha\ and K-line cores. G:
    spicules.  [\dots] The distance between the bushes is 30\,000\,km.
    Taken from \citetads{1967IAUS...28..293N}, % Noyes
    including this caption.}
\end{figure}
%===========================================================================

%===========================================================================
\begin{figure}
%RR small version for Astroph
  \centerline{\includegraphics[width=\textwidth]{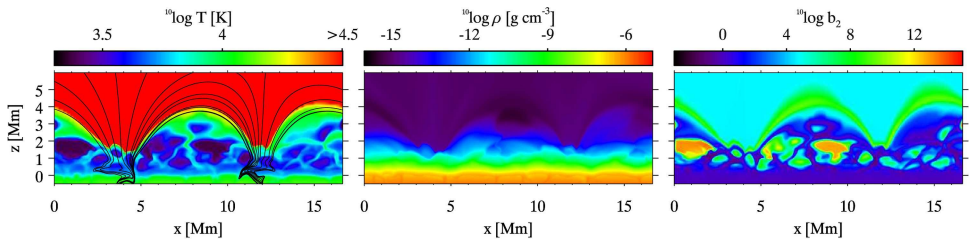}}
  \caption[]{\label{fig:hion2}%
    Cross-sections through a snapshot of a time-dependent 2D MHD
    simulation.
    The computational domain includes the top of the convection zone
    and reaches up to the corona. First panel: temperature, with
    superimposed field lines that are selected to chart the two
    magnetic concentrations.  Second panel: gas density.  Third panel:
    NLTE overpopulation of the $n\is2$ level of hydrogen, setting the
    \Halpha\ opacity.  A movie, available at
    \url{http://www.astro.uu.nl/~rutten/rrweb/rjr-movies}, of the
    temporal variation of these cross-sections demonstrates that the
    narrow blue-green fronts in the clapotisphere under the
    internetwork canopy in the first panel represent shocks that
    travel upwards, mostly at a slant and with much mutual interference.
    They delimit very cool clouds (blue-black in the first panel) with
    gigantic \HI\ $n\is2$ overpopulation (green-orange in the third
    panel).  Taken from
    \citetads{2007A&A...473..625L}. %C Leenaarts hion2
  }
\end{figure}
%===========================================================================

{\em Photosphere\/} is a better descriptor than `surface' for the thin
near-spherical shell where the visible and infrared solar continua
originate.  It extends a few hundred km from the $\tau_{500} \is 1$
Eddington-Barbier depth at which the radially emergent continuous
intensity at $\lambda \is 500$\,nm is approximately given by the
continuum source function, which is dominated by
bound-free \Hmin\ transitions.  These contribute so much opacity that
sunlight escapes only at much lower atmospheric gas density than that
of the transparent air surrounding us.  Slightly deeper continuum
escape occurs in the opacity minima near $\lambda \is 400$\,nm (with
shortward onset of important metal ionisation edges that provide the
\Hmin\ electrons) and $\lambda \is 1.6\,\mu$m (\Hmin\ bound-free
threshold, with longward increase of the \Hmin\ free-free
contribution).

Yet deeper escape, about 100--200\,km, takes place within slender
magnetic concentrations of kilogauss strength. In these, the magnetic
pressure contribution to hydrostatic balancing reduces the gas
density.  Such fluxtubes constitute network as loose `filigree'
alignments along intergranular lanes at supergranular boundaries.  At
larger activity, they constitute plage in the form of denser clusters
that inhibit normal granulation.  

In addition, there appears to be abundant magnetism at much lower
strength.  Other quiet-photosphere ingredients are the ubiquitous
granulation due to turbulent convection, overshoot phenomena including
internal gravity waves, the supergranulation, and copious acoustics
that are dominated by global $p$-mode standing-wave interference but
also contain outward propagating waves from local excitation at
granular scales.

{\em Chromosphere\/} does not stand for a spherical shell but for a thin,
very warped and highly dynamic interface surface that comes down deep
in and near kilogauss concentrations but rides high on acoustic shocks
in otherwise cool internetwork (cell interior) gas.  I call the
latter domain the `clapotisphere'
(\citeads{1995ESASP.376a.151R}) % Rutten Asilomar review
and reserve `chromosphere' for the fibrilar
% \footnote{ Using `fibril'
%   as a generic name for slender \Halpha\ features.  Traditionally,
%   shorter near-network \Halpha\ features making up `rosettes' are
%   called mottles, longer ones crossing cell interiors
%   fibrils.}  
canopies seen in \Halpha.  These appear to be structured by the fields
that extend from network and plage.  The fibrils correspond to the
off-limb spicule forest whose pink Balmer-line emission gave the
chromosphere its name
(\cite{1868RSPS...17..131L}). % Lockyer chromoshere
%% ; see \citeads{2010MmSAI..81..565R}). % SacPeak review
The transition region to the corona is likely a thin envelope to the
fibrilar chromosphere.
%RR upper or sheath? that's the question
The principal ingredients defining chromospheric structure and
dynamics are, for decreasing activity, magnetic reconnection, current
heating, %RR like Juan spicule-II simulation
Alfv\'en waves, magnetically guided and/or converted acoustic waves,
possibly gravity waves and torsional waves, and photon losses
in strong lines.

In terms of physics, the principal quiet-Sun photospheric agents are
gas dynamics and near-LTE radiation loss outside magnetic
concentrations, magnetohydrodynamics (MHD) within the latter.  These
processes are presently emulated well in 3D time-dependent simulations
of photospheric fine structure.  The spatial simulation extent is
still too small to contain full-fledged active regions, but sunspots
% (\citeauthor{2009Sci...325..171R} % Rempel penumbra pair
% \citeyearads{2009Sci...325..171R}, % Rempel penumbra pair
% \citeyearads{2009ApJ...691..640R}; % Rempel++ MHD spot
(\eg\ \citeads{2011ApJ...740...15R}) % Rempel deeper spot
and supergranules
(\citeads{2008AIPC.1043..234U}; % Ustyugov supergranulation
\citeads{2009ASPC..416..421S}) % Stein++ supergranulation
come into reach.  Higher up, the radiation losses become severely
non-equilibrium (NLTE, PRD, time-dependent population rates) and the
magnetogasdynamics becomes multi-fluid.  These complexities constitute
a challenging but promising modelling frontier.

%%%%%%%%%%%%%%%%%%%%%%%%%%%%%%%%%%%%%%%%%%%%%%%%%%%%%%%%%%%%%%%%%%%%%%%%%%%%
\section{The scene} \label{sec:scene}
%%%%%%%%%%%%%%%%%%%%%%%%%%%%%%%%%%%%%%%%%%%%%%%%%%%%%%%%%%%%%%%%%%%%%%%%%%%%

Figure~\ref{fig:Noyes} sets the quiet-Sun scene for this overview.
This sketch comes from the outstanding discussion by
\citetads{1967IAUS...28..293N} % Noyes
of the ingenious Doppler imaging first described in the seminal
`Preliminary report' of
\citetads{1962ApJ...135..474L}. % Leighton+Simon+Noyes I
The latter contained the discovery of the supergranulation, reversed
granulation, five-minute oscillation, upward-propagating chromospheric
waves, and rapidly changing chromospheric flows, all in a single
paper!  The various symbols in Noyes's sketch define characteristic
fine structure of the solar photosphere and chromosphere.  The
photospheric granulation, waves, and strong-field concentrations are
now largely understood\footnote{Understood in terms of their
  structural physics and how their observational diagnostics arise.
  The unsolved riddles of how network fields come, assemble, cancel,
  disperse, and go are outside the scope of this article, as are {\em
    a fortiori\/} the structure, formation, evolution, and decay of
  active regions and filaments and their outbursts. The dynamo and
  activity cycle remain grand questions (\cf\
  \citeads{2011sswh.book...39S}).} % Spruit critique dynamo
The photospheric supergranulation and the chromospheric structures
and flows are not.

Figure~\ref{fig:hion2} from
\citetads{2007A&A...473..625L} %C Leenaarts hion2
is not a cartoon but a snapshot from a time-dependent simulation
representing state-of-the-art numerical implementation of the physics
of MHD and radiative transfer in the solar atmosphere.  The code of
\citetads{2007ASPC..368..107H} % Hansteen++ Coimbra stagger code
was used; other codes are described by
\citetads{2002AN....323..213F}, % Freytag+Steffen+Dorch 1st COBOLD
\citetads{2006ASPC..354..345S}, % Schaffenberger++ MHD sim
\citetads{2005A&A...429..335V}, % Voegler++ MURAM
\citetads{2005ApJ...618.1020G} % Gudiksen+Nordlund MHD
and
\citetads{2011A&A...531A.154G}. % Gudiksen++ Bifrost
In this case, one spatial dimension was sacrificed to make
non-equilibrium evaluation of hydrogen population rates tractable
(\cf\ \citeads{1976A&A....47...65K}; % Kneer+Nakagawa time-dependent
\citeauthor{1992ApJ...397L..59C} % Carlsson+Stein 1D time-dependent
\citeyearads{1992ApJ...397L..59C}, % Carlsson+Stein 1D time-dependent
\citeyearads{2002ApJ...572..626C}; % Carlsson+Stein Hion time-dependent
\citeads{2003ApJ...589..988R}). % Rammacher+Ulmschnieder time-dependent
This is essential above the photosphere where shocks abound.  The
slowness of H ionisation/recombination balancing in the cool
post-shock aftermaths makes hydrogen a much less effective internal
energy buffer than it would be for instantaneous statistical
equilibrium or LTE, and strongly affects the thermodynamics.  The slow
post-shock balancing also causes huge NLTE over-opacities of \Halpha\
(third panel).

The two figures have much similarity.  Each contains granules,
oscillations, flows, and two magnetic concentrations in the
photosphere (unipolar in figure~\ref{fig:Noyes}, bipolar in
figure~\ref{fig:hion2})
%%, with different horizontal separation)
whose fields spread out at larger height.

There are also dissimilarities.  A key one concerns the `chromospheric
flow field seen in \Halpha' marked by D in figure~\ref{fig:Noyes} that
describes flows along cell-covering fibrils.  In \Halpha\ filtergrams
such long internetwork fibrils appear ubiquitously, constituting an
opaque chromospheric blanket all over the Sun except in extremely
quiet areas.  At the limb these fibril canopies provide an opaque
floor to the \Halpha\ spicule forest, \ie\ much of Lockyer's
chromosphere.  However, the simulation snapshot
(figure~\ref{fig:hion2}) does not contain such internetwork
fibrils. The green arches in the third panel look suggestively as
such, but these remain optically thin in \Halpha\ even while marking
\Halpha\ opacities as much as $10^8$ in excess of LTE.

Underneath these fibrils the sketch specifies wave motions;
correspondingly, under the arches the simulation snapshot has large
clapotispheric clouds of cool gas that are permeated by repetitive
shocks.  The latter push the chromospheric interface with the corona
up to heights around 3--4\,Mm, as high as the flow-mapping \Halpha\
fibrils in the sketch.  Near the magnetic concentrations the shocks
are field-guided and jut out with repetitive extension and retraction
to 2--3\,Mm height (figure~3 of
\citeads{2007A&A...473..625L}). %C Leenaarts hion2.
More on these dynamic fibrils below, and also on
straws\,/\,spicules-II\,/\,RBEs and weak fields which do not figure in these
diagrams.

%%%%%%%%%%%%%%%%%%%%%%%%%%%%%%%%%%%%%%%%%%%%%%%%%%%%%%%%%%%%%%%%%%%%%%%%%%%%
\section{Photosphere} 
%%%%%%%%%%%%%%%%%%%%%%%%%%%%%%%%%%%%%%%%%%%%%%%%%%%%%%%%%%%%%%%%%%%%%%%%%%%%

%%%%%%%%%%%%%%%%%%%%%%%%%%%%%%%%
\rrsubsection{Overall structure}
%%%%%%%%%%%%%%%%%%%%%%%%%%%%%%%%

\topic{1D modelling}
%%%%%%%%%%%%%%%%%%%%
\citet{1906WisGo.195...41S} % Schwarzschild with me as "author"
wrote,
in Loeser's translation in the compilation by
\citetads{1962sppp.book.....M}: % Menzel selected RT papers
{ \em ``the sun's surface displays changing conditions and stormy
  variations [\ldots] It is customary, as first approximation, to
  substitute mean steady-state conditions for these spatial and
  temporal variations, thus obtaining a mechanical or `hydrostatic
    equilibrium' of the solar atmosphere''} and established that
{\em ``the equilibrium conditions of the solar atmosphere correspond
  generally to those of radiative equilibrium''.}  In cool-star abundance
studies, stellar-atmosphere modellers usually assume these equilibria
plus mixing-length convection at the bottom of the photosphere.
% combined with LTE radiative transfer
%  as in Kurucz' ATLAS codes (\eg\ %
% \citeyearads{1974SoPh...34...17K}, % Kurucz
% \citeyearads{1979ApJS...40....1K}; % Kurucz
% \citeads{2009A&A...508..401C}) and the Uppsala MARCS code (\eg\
% \citeads{1975A&A....42..407G}; % Gustaffson++ grid
% %BB \citeads{1976A&AS...23...37B}; % BELLEA
% %BB \citeads{2008A&A...486..951G}; % Gustafsson++ MARCS grid
% \citeads{2010arXiv1011.2092V}), % S star grid
% or with NLTE radiative transfer as in MULTI (\eg\
% \citeads{1986UppOR..33.....C}; % Carlsson MULTI
% \citeads{2009A&A...500.1221F}), % Fabbian++ oxygen stars
% TLUSTY (\eg\ \citeads{1988CoPhC..52..103H}; % Hubeny: TLUSTY
% \citeads{1995ApJ...439..875H}; % Hubeny+Lanz TLUSTY
% \citeads{2010AIPC.1268...73H}), % Hubeny school
% or PHOENIX (\eg\ \citeads{1994ApJ...422..831H}; % Hauschildt++ PHOENIX
% \citeads{2010A&A...522A.102S}). % Seelmann++ Phoenix 3D

In contrast, solar abundance determiners have typically used empirical
spectrum-fitting models, in particular the HOLMUL update by
\citetads{1974SoPh...39...19H} % HOLMUL
of the LTE line-fitting model of
\citetads{1967ZA.....65..365H}. % Holweger model
They preferred this over the much more sophisticated NLTE
continuum-fitting FALC model of
\citetads{1993ApJ...406..319F} % FAL including FALC
because HOLMUL has no chromospheric temperature rise which would
produce self-reversals in LTE-computed strong lines that are not
observed.  The mid-photosphere parts of these empirical models
are so similar and so close to radiative-equilibrium predictions that
LTE and RE seem reasonable first approximations where
\Hmin\ radiation loss is the major stratification agent.
% It is likely that the actual photosphere obeys LTE-RE equilibria even in
% and near magnetic concentrations, but dynamically in small-scale 3D
% geometry including hot-wall radiation loss and irradiation.

\topic{NLTE masking}
%%%%%%%%%%%%%%%%%%%%
FALC's classic VAL3C predecessor, formulated in the three monumental
papers of Vernazza, Avrett \& Loeser
(\citeyearads{1973ApJ...184..605V}, % VALI
\citeyearads{1976ApJS...30....1V}, % VALII
\citeyearads{1981ApJS...45..635V}), % VALIII
had a significantly cooler upper photosphere than HOLMUL.  Such a
steeper temperature gradient produces deeper lines, but also
appreciable near-ultraviolet overionisation of species such as \FeI\
(\citeads{1972PhDT.........7L}). % Lites thesis
The resulting NLTE reduction of \FeI\ opacities offsets the line
strengthening; \citetads{1982A&A...115..104R} % Rutten+Kostik
called this fortuitous cancellation `NLTE masking'.  Avrett's
subsequent and most recent 
% (VAL3C$^\prime$ in
% \citeads{1985cdm..proc...67A}, % Avrett Keil workshop
% MACCKL in \citeads{1986ApJ...306..284M}) % MACCKL
(\citeads{2008ApJS..175..229A}) % Avrett+Loeser update SUMER
modelling includes a large number of ultraviolet lines to provide a
quasi-continuous line haze 
%% (\citeads{1975SoPh...45..115Z}; % Zwaan line-haze model sunspots
(\eg\ \citeads{1980A&A....90..239G}; % Greve+Zwaan UV haze
\citeads{2009AIPC.1171...43K}) % Kurucz line list
that resulted in a less steep, HOLMUL-like photosphere which was
copied into FALC.  All these lines are set to share an ad-hoc
transition from LTE to pure scattering to avoid core reversals (see
\citeads{1988ASSL..138..185R}). % Rutten Capri review
This NLTE opacity issue crops up again in modelling photospheric fine
structure with steep
% as well as shallow 
temperature gradients
(\citeads{2001ApJ...550..970S}). % Shchukina FeI NLTE

\topic{Microturbulence}
%%%%%%%%%%%%%%%%%%%%%%%
All 1D modelling uses Struve's microturbulence as a free adjustment
factor to account for ignored fine structure in matching observed
spectral lines.  It is the most arbitrary of the four classical
stellar-atmosphere parameters (effective temperature, surface gravity,
metallicity, turbulence) --
\citetads{1949ApJ...110..329C} % Chandrasekhar turbulence
quoted Russell's attribution {\em ``to the direct intervention of the
Deity''\/}.  In 1D solar modelling it became a bag of fudge
parameters. Macroturbulence and height dependence were added,
\citetads{1977SoPh...54..313D} % CdeJ+Vermue filters
defined functional kinetic-energy spectrum filters,
\citetads{1977ApJ...218..530G} % Gray
added anisotropy.  VAL3C and FALC shared sizable microturbulence
with large variation with height (figure~11 of
\citeads{1981ApJS...45..635V}). % VAL3
The underlying assumption that the unresolved fine structure and
dynamics can be described as a Gaussian convolution (of the extinction
coefficient for micro, of the emergent intensity profile for macro,
respectively) appears untenable
(\citeads{2011ApJ...736...69U}). % Uitenbroek+Criscuoli 1D-3D

%%%%%%%%%%%%%%%%%%%%%%%%%%%%%
\rrsubsection{Fine structure}
%%%%%%%%%%%%%%%%%%%%%%%%%%%%%

\topic{Granulation}
%%%%%%%%%%%%%%%%%%%
The granulation became initially understood by Nordlund's early
simulations (\eg\ Nordlund
\citeyearads{1982A&A...107....1N}, % Nordlund granulation I
\citeyearads{1984ssdp.conf..181N}) % Nordlund Keil workshop
and more so when he teamed up with Stein (\eg\ Stein \& Nordlund
\citeyearads{1989ApJ...342L..95S}, % Stein+Nordlund topology convection
\citeyearads{1998ApJ...499..914S}; % Stein+Nordlund general properties
\citeads{1990CoPhC..59..119N}; % Nordlund+Stein
\citeads{2009LRSP....6....2N}) % Nordlund+Stein+Asplund review
and with the identification of radiative surface cooling as the major
driver
(\citeads{1995ApJ...443..863R}). % Rast exploders make driving downflows
Current granulation research targets small vortices 
(\citeauthor{2008ApJ...687L.131B}
\citeyearads{2008ApJ...687L.131B}, % Bonet++ vortices
\citeyearads{2010ApJ...723L.139B}; % Bonet++ Sunrise vortices
\citeads{2010ApJ...723L.180S}; % Steiner++ Sunrise vortices
\citeads{2011ApJ...727L..50K}; % Kitiashvili++ sim vortices
\citeads{2011A&A...526A...5S}), % Shelyag++ sim vortices
supersonic flows
(\citeads{2009ApJ...700..284B}; % Bellot supersonic granular flows
\citeads{2011A&A...532A.110V}), % Vitas++ supersonic granular flows
flow bending
(\citeads{2010ApJ...723L.159K}), % Khomenko++ granule flow bends
and intergranular jets (\citeads{2010ApJ...714L..31G}; % Goode++
\citeads{2011ApJ...736L..35Y}). % Yurchyshyn++ DST
% discovered with the new NST\footnote{New Solar
%    Telescope: \url{http://www.bbso.njit.edu/nst_project.html}.}.

\topic{Abundances}
%%%%%%%%%%%%%%%%%%
The use of 3D Nordlund-Stein granulation simulations
instead of 1D modelling with ad-hoc turbulence
% by Asplund and coworkers 
resulted in substantial downward revisions of key abundances,
upsetting helioseismology and stellar evolution theory (\eg\ 
\citeauthor{2004A&A...417..751A}
\citeyearads{2004A&A...417..751A}, % Asplund++ Oxygen
\citeyearads{2005A&A...435..339A}; % Asplund++ O abundance ++
\citeauthor{2009A&A...507..417P}  % Pereira++ Oxygen abundance
\citeyearads{2009A&A...507..417P}, % Pereira++ Oxygen abundance
\citeyearads{2009A&A...508.1403P}; % Pereira++ Oxygen abundance
\citeads{2009LRSP....6....2N}, % Nordlund+Stein+Asplund
\citeads{2009ARA&A..47..481A}). % Asplund++ abundances
However, 
\citeauthor{2009MmSAI..80..643C}
(\citeyearads{2009MmSAI..80..643C}, % Caffau++ abundances
\citeyearads{2011SoPh..268..255C}) % Caffau++ abundances
% using the CO5BOLD 3D code of
% \citetads{2002AN....323..213F}, % Freytag+Steffen+Dorch 1st COBOLD
report only small changes from the 1D to 3D paradigm shift and do not
confirm such drastic revision.
\citetads{2010ApJ...724.1536F} % Fabbian++ MHD abundance
found 
% with the 3D magnetoconvection code of
% \citetads{2006ApJ...642.1246S} % Stein+Nordlund magnetoconvection
that adding moderate magnetic flux has appreciable influence on iron
abundance determination.  These issues remain open.

\topic{Reversed granulation}
%%%%%%%%%%%%%%%%%%%%%%%%%%%%
In the mid-photosphere the granular intensity contrast reverses.  The
observations of
\citetads{2004A&A...416..333R} % Rutten++ tomo2 reversed granulation
were well reproduced by the
% CO5BOLD 
simulation of
\citetads{2005A&A...431..687L}. % Leenaarts+Wedemeyer tomo3
\citetads{2007A&A...461.1163C} % Cheung++ reversed granulation
provided detailed explanation using the MHD code of
\citetads{2005A&A...429..335V}. % Voegler++ MURAM
The phenomenon comes from overturning flows and produces marked
imaging asymmetry between the wings of \NaI\ \Done\
(\citeads{2011A&A...531A..17R}). % Rutten++ NaD1 revgran

\topic{Acoustic oscillations}
%%%%%%%%%%%%%%%%%%%%%%%%%%%%%
The simulations have self-excited acoustic box modes that
are not too dissimilar from global $p$-modes
(\citeads{2001ApJ...546..576N}; % Nordlund+Stein box modes as p-modes
\citeads{2001ApJ...546..585S}). % Stein+Nordlund box modes as p-modes
They serve to study stochastic mode excitation in near-surface
convection for the Sun and other stars (\eg\ 
\citeauthor{2003A&A...403..303S}
\citeyearads{2003A&A...403..303S}, % Samadi improve exc model
\citeyearads{2007A&A...463..297S}). % Samadi stochastic osc exc across HR
Earlier, Goode and coworkers searched for identifiable $p$-mode
exciters (\eg\
\citeads{1993ApJ...408L..57R}; % Restaino+Stebbins+Goode, wave excitation
\citeads{1995ApJ...444L.119R}; % Rimmele+Goode+Harold+Stebbins
\citeads{1998ApJ...495L..27G}; % Goode+Strous+Rimmele+Stebbins
\citeads{2000ApJ...535.1000S}) % Strous+Goode+Rimmele, events dynamics
for which the collapsars (small vanishing granules) of
\citetads{1999ApJ...524..462R} % Rast, thermal plume = source
and
\citetads{2000ApJ...541..468S} %C Skartlien+Stein+Nordlund, waves collapsar
became the principal candidate.  These indeed generate
excess acoustics %% but not one-to-one
(\citeads{2002A&A...390..681H}). % Hoekzema+Rimmele+Rutten fluxca

\topic{Magnetic concentrations}
%%%%%%%%%%%%%%%%%%%%%%%%%%%%%%%
Figure~\ref{fig:Noyes} is from before the
identification of magnetic network elements as kilogauss fluxtubes by
\citetads{1978A&A....70..789F} % Frazier+Stenflo kilogauss
with Stenflo's (\citeyearads{1973SoPh...32...41S}) % Stenflo line ratio
line ratio technique.
% (\cf\
% \citeads{2010mcia.conf..101S}). % Stenflo Evershed conf
In the magnetostatic thin-fluxtube model of
\citetads{1976SoPh...50..269S} %? Spruit, magnetostatic fluxtube
following ideas of
\citetads{1967SoPh....1..478Z}, %C Zwaan: invisible sunspots
the inside magnetic pressure balances part of the outside gas
pressure; the latter's outward drop makes the fluxtube expand
with height.
This paradigm was put on firm observational footing by Solanki and
coworkers through best-fit modelling of the spatially unresolved
signature of small kilogauss concentrations in multi-line
spectropolarimetry
% (\eg\
% \citeads{1985A&A...148..123S}; % Solanki+Stenflo: fluxtubes from Fe lines
% \citeads{1986A&A...168..311S}; % Solanki: fluxtubes with velocities
% Solanki \etal\
% \citeyearads{1988A&A...189..243S},%C Solanki+Steenbock netw1,netw2,plage
% \citeyearads{1992A&A...262L..29S}; % Solanki+Brigljevic fluxtube model
% \citeads{1993A&A...268..736B}; %T Bunte+Solank+Steiner, wine glass model
(see the extensive review by
\citeads{1993SSRv...63....1S}), %C Solanki new testament
and subsequently confirmed with MHD simulations, first in 2D (\eg\
\citeauthor{1994A&A...285..648G} % GrossmanDoerth++: deep layers fluxtubes
\citeyearads{1994A&A...285..648G}, % GrossmanDoerth++: deep layers fluxtubes
\citeyearads{1998A&A...337..928G}; % GrossmanDoerth++, collapse
\citeads{1998ApJ...495..468S}; % Steiner++, Freiburg fluxsheet
\citeads{2001SoPh..203....1G}; % Gadun++ 2D fluxtubes
see also the two examples in figure~\ref{fig:hion2}), then in 3D (\eg\
\citeads{2004ApJ...610L.137C}; % Carlsson++ G-band simulation
\citeads{2005A&A...429..335V}; % Voegler++ MURAM MCs
\citeads{2009A&A...504..595Y}). % Yelles++ check fluxtube from 3D MHD
Newer research addresses wave excitation and diagnostics
(\citeads{2008SoPh..251..589K}; % Khomenko++ fluxtube waves
\citeauthor{2009A&A...508..951V}
\citeyearads{2009A&A...508..951V}, % Vigeesh++ fluxtube wave
\citeyearads{2011SoPh..tmp..349V}; % Vigeesh++ fluxtube wave diagnostics
\citeads{2011ApJ...730L..24K}). % Kato++ magnetic pumping

\topic{Magnetic bright points}
%%%%%%%%%%%%%%%%%%%%%%%%%%%%%%
Many studies of small strong-field concentrations employ their alter-ego as
network bright points (or magnetic knots, filigree grains, facular
points, \etc) which are easier to observe, easiest in the G~band
of CH lines around $\lambda\is430.5$~nm following
\citetads{1984SoPh...94...33M} % Muller+Roudier G-band BP's
but also in the wings of strong lines
(\citeads{2006A&A...452L..15L}). % Leenaarts++ BP diagnostics
They are not points, becoming flowers and ribbons with much
substructure at high
resolution (\citeads{2004A&A...428..613B}; % Berger++, SST flowers
% and breaking up at NST resolution
\citeads{2010ApJ...725L.101A}). % Abramenko++ BPs with NST
%RR seems lower to me
The upshot of the extended literature reviewed by
\citetads{2009SSRv..144..275D} % de Wijn++ small-scale fields
is that they are useful indicators of magnetic concentrations in
imaging `proxy' magnetometry, but without precise one-to-one
correspondence to the actual fields.  Their brightness was explained
by \citetads{1976SoPh...50..269S} % Spruit hot wall heating
and
\citetads{1981SoPh...70..207S}. % Spruit+Zwaan fluxtube size-brightness
It is not due to magnetic energy dissipation but comes from the hot
walls that fluxtubes have below the outside surface because the
viewing depth inside is 100--200~km deeper from the partial
evacuation.  In molecular bands dissociation increases the contrast
with the outside granulation, as does ionisation for atomic lines and
smaller collisional damping for strong-line wings.  Towards the limb
facular brightening results because viewing along a slanted line of
sight penetrates through the tube into the granule behind it.
Detailed MHD simulations give good bright-point and facular agreement
with observations (\eg\
\citeads{2004ApJ...607L..59K}; % Keller++: faculae
\citeads{2004A&A...427..335S}; % Shelyag++ G-band simulation
\citeads{2004ApJ...610L.137C}; % Carlsson G-band BPs
\citeads{2005A&A...430..691S}; % Steiner faculae
\citeads{2006A&A...449.1209L}; % Leenaarts++ BPs blue wing Ha
\citeads{2009A&A...499..301V}; % Vitas++ MnI
%% except possibly the brightness contrast  
%% (\citeads{2007ApJ...668..586U}, % Uitenbroek++ Gband contrast
  %RR too simple on straylight
\citeads{2009A&A...503..225W}). % Wedemeyer+Rouppe contrasts
  %RR better straylight treatment
Irradiance modelling 
% (\eg\ \citeads{1999A&A...345..635U}) % Unruh++ irradiance modeling
must cope with the inherently 3D nature of limbward facular brightening
(\citeads{2009AIPC.1094..768U}). % Unruh++ simulation vs SATIRE

\topic{Weak and horizontal fields}
%%%%%%%%%%%%%%%%%%%%%%%%%%%%%%%%%%
Neither figure~\ref{fig:Noyes} nor figure~\ref{fig:hion2} shows the
weak tangled internetwork fields detected with sensitive 
spectropolarimetry
(\citeads{2003A&A...408.1115K}; % Khomenko++ IN fields
\citeads{2004Natur.430..326T}; % JTB+Shchukina weak fields
\citeads{2007PASJ...59S.571L}; % Lites++ Hinode weak+hor
\citeads{2010ApJ...714L..94M}) % Martinez-Gonzalex++ little omega loops
and predicted by near-surface convective dynamo simulations
(\citeads{1999ApJ...515L..39C}; % Cattaneo surface dynamo
\citeads{2007A&A...465L..43V}). % Voegler+Schuessler surface dynamo
%% \citeads{2007ApJ...665.1469A}) % Abbett convection dynamo
They seem to close on granular scales in the mid-photosphere,
appearing there as relatively strong horizontal field
(\citeads{2008ApJ...672.1237L}; % Lites++ hor fields
\citeads{2008ApJ...680L..85S}; % Steiner++ hor field sim
\citeads{2008A&A...481L...5S}). % Voegler++ hor field sim

\topic{Inversion modelling}
%%%%%%%%%%%%%%%%%%%%%%%%%%%
The La Laguna inversion school initiated by Ruiz-Cobo et al.\
(\citeyearads{1990Ap&SS.170..113R}, % Ruiz Cobo start SIR
\citeyearads{1992ApJ...398..375R}) % SIR
has produced multiple codes that emulate HOLMUL-like empirical model
construction by best-fit stratification modelling per observed pixel
(\eg\ \citeauthor{2001ApJ...553..949S}
\citeyearads{2001ApJ...553..949S}, % Socas++ LILIA
\citeyearads{2008ApJ...674..596S}; % Socas++ multi-line inversions
\citeads{2008ApJ...683..542A}; % Asensio+JTB HAZEL
\citeads{2010MmSAI..81..716D}; % DelaCruz 8542+NICOLE
\citeads{2010SoPh..tmp..254B}). % Borrero++ VSIV inversion HMI
They serve especially to map flows and magnetic fields from
full-Stokes line profiles.  Many use the Milne-Eddington approximation
of constant line/continuum opacity ratio; some use slab geometry.
Many fit spline functions with height through a few anchor points.
The approach generally works well in the photosphere where most radial
behaviour is smooth.

%%%%%%%%%%%%%%%%%%%%%%%%%%%%%%%%%%%%%%%%%%%%%%%%%%%%%%%%%%%%%%%%%%%%%%%%%%%%
\section{Clapotisphere} 
%%%%%%%%%%%%%%%%%%%%%%%%%%%%%%%%%%%%%%%%%%%%%%%%%%%%%%%%%%%%%%%%%%%%%%%%%%%%

%%%%%%%%%%%%%%%%%%%%%%%%%%%%%%%%
\rrsubsection{Overall structure}
%%%%%%%%%%%%%%%%%%%%%%%%%%%%%%%%

\topic{1D modelling}
%%%%%%%%%%%%%%%%%%%%
The six VAL3 models of \citetads{1981ApJS...45..635V} % VALIII
representing quiet-Sun conditions were made by fitting bins of
different observed ultraviolet intensities with different temperature
stratifications, all sharing the basic structure of photospheric
decline, temperature minimum, extended raised chromospheric plateau,
and a steep rise to coronal values called the transition region.
Brighter spectra were reproduced by models that are slightly hotter
throughout their upper parts and have a deeper-located transition
region.  With these models the term `chromosphere' became to mean the
raised-temperature plateau between the VAL3C temperature minimum at
$h\is 500$\,km and transition region at $h\is2100$\,km, rather than
\Halpha\ fine structure.

However, collapsing the internetwork parts and network parts,
respectively, of figures~\ref{fig:Noyes} and \ref{fig:hion2} to column
averages would make sense only if the spatiotemporal fluctuations were
small, but they are not at heights where the atmosphere suffers
frequent shocks.  The slight differences between the different VAL3
models (and for the similar grids of
\citeauthor{1993ApJ...406..319F} % FAL3 including FALC
(\citeyearads{1993ApJ...406..319F}, % FAL3 including FALC
\citeyearads{2002ApJ...572..636F}, % FAL4 = H and He mass flows
\citeyearads{2006ApJ...639..441F}, % Fontenla++ moderate models I
\citeyearads{2007ApJ...667.1243F}, % Fontenla++ moderate models II
\citeyearads{2009ApJ...707..482F}) % Fontenla++ moderate models III
should be contrasted with the much larger temporal variations along
columns in figure~\ref{fig:hion2}.  In a
movie\footnote{\url{http://www.astro.uu.nl/~rutten/rrweb/rjr-movies/hion2_fig2_movie.mov}}
of this behaviour, the internetwork clapotisphere consists of very
cool gas that is every few minutes ridden through by shocks.  That
such shocks invalidate static modelling was established already
with the beautiful \CaII\ \HtwoV\ grain simulation by
\citeauthor{1997ApJ...481..500C}
(\citeyearads{1994chdy.conf...47C}, % Carlsson+Stein miniworkshop
\citeyearads{1995ApJ...440L..29C}, % Carlsson+Stein, does chrom T-rise exist?
\citeyearads{1997ApJ...481..500C}). % Carlsson+Stein, H2v grain simulation

The existence of cool clouds 
%% (`COmosphere') 
that are incompatible with static 1D modelling was long before
advocated by Ayres on the basis of deep CO line cores (\eg\
\citeads{1981ApJ...244.1064A}; % Ayres bifurcation
\citeads{1996ApJ...460.1042A}; % Ayres+Rabin CO revisited
\citeads{2002ApJ...575.1104A}). % Ayres fulltime comosphere
In simulations the clapotispheric gas gets as cool as 2000\,K.
\citetads{2011A&A...530A.124L} % Leenaarts+Carlsson cool
conclude that such low temperatures are unavoidable unless there is
weak-field heating not present in such simulations.

ALMA may deliver direct mm-radiation diagnostics of the spatiotemporal
temperature behaviour (\eg\
\citeauthor{2004A&A...419..747L} % Loukitcheva++ mm fluctuations
\citeyearads{2004A&A...419..747L}, % Loukitcheva++ mm fluctuations
\citeyearads{2006A&A...456..713L}.) % Loukitcheva++ mm fluctuations

%%%%%%%%%%%%%%%%%%%%%%%%%%%%%
\rrsubsection{Fine structure}
%%%%%%%%%%%%%%%%%%%%%%%%%%%%%

\topic{Acoustic shocks}
%%%%%%%%%%%%%%%%%%%%%%%
Many shocks in the movie companion to Figure~\ref{fig:hion2} run at a
slant, which explains why
\citetads{1997ApJ...481..500C} %C Carlsson+Stein, H2v grain simulation
found only a few locations in the observations of
\citetads{1993ApJ...414..345L} % Lites+Rutten+Kalkofen, network dynamics
with well-defined \HtwoV\ grain sequences that they could faithfully
reproduce one-dimensionally.  The grains obtain their characteristic
violet/red asymmetry from the presence of both a hot shock and
higher-up post-shock downdraft along the line of sight; grain
sequences imply vertical shock propagation.  The slanted-shock
interaction patterns indeed resemble clapotis (wild wave interference
on rivers and oceans).

\topic{Gravity waves}
%%%%%%%%%%%%%%%%%%%%%
Gravity waves should be copiously excited in the convective overshoot
above the granulation (\eg\
\citeads{1963ApJ...137..914W}; % Whitaker coronal heating grav waves
\citeads{1967SoPh....2..385S}; % Stein
\citeads{1967IAUS...28..429L}; % Lighthill corrected by me 
\citeauthor{1981ApJ...249..349M} % Mihalas+Toomre I
\citeyearads{1981ApJ...249..349M}, % Mihalas+Toomre I
\citeyearads{1982ApJ...263..386M}), % Mihalas+Toomre II
but are hard to diagnose (\eg\
\citeads{1980ApJ...236L.169B}; % Brown+Harrison from brightness
\citeads{1989A&A...213..423D}; % Deubner+Fleck I
\citeads{2003A&A...407..735R}). % Rutten+Krijger TRACE II grav waves
Recent estimates combining Fourier observations with simulations,
but applying linear theory, suggest that the gravity-wave energy flux into
the upper chromosphere is comparable to the acoustic flux
(\citeads{2008ApJ...681L.125S}; % Straus++ grav waves
\citeads{2011A&A...532A.111K}). % Kneer+Bello grav wave flux
These waves remain understudied.

\topic{Quietest Sun}
%%%%%%%%%%%%%%%%%%%%
In the quietest areas \Halpha\ line-centre images show no cover-all
fibril carpet but only network rosettes consisting of short mottles,
and very dynamic cell interior between these.  A fast-cadence \Halpha\
movie\footnote{%
\url{http://www.astro.uu.nl/~rutten/rrweb/rjr-movies/2006-06-18-quiet-ca+hawr.avi}}
of the latter from the data of
\citetads{2007ApJ...660L.169R} % Rouppe+BdeP quiet fibrils
shows mushrooming three-minute oscillation blobs that break up into
very narrow, fast-moving, probably hot, filamentary structures.
Possibly these mark shock interference, or conversion into weak-field
canopy waves, or tiny magnetic filaments as described by
\citetads{2006ASPC..354..345S}, % Schaffenberger++ MHD sim
or shock-excited high-frequency oscillations as proposed by
\citetads{2008ApJ...683L.207R}. % Reardon++ turbulence from buffeting
\citetads{2009A&A...507L...9W} % Wedemeyer+Rouppe swirls
reported swirl motions in such areas.  These quiet-Sun phenomena also
need further study, but require exceedingly high spatial and temporal
resolution.

\topic{Wave-field interaction}
%%%%%%%%%%%%%%%%%%%%%%%%%%%%%%
The shocks push the clapotispheric transition region to large height
because magnetic refraction and mode conversion become important only
at low plasma-beta
(\citeads{Bogdan+others2002}; % Potsdam Thinkshop, mix+interfere
\citeads{2002ApJ...564..508R}). % Rosenthal++ waves and field
The ubiquitous presence of shock waves in both the observations and
MHD simulations suggests that weak fields play no large role in the
clapotisphere (\cf\ \citeads{2010mcia.conf..166S}), % Steiner Evershed
nor the multi-scale magnetic carpet proposed by
\citetads{2003ApJ...597L.165S}. % Schrijver+Title IN carpet
Gravity waves convert easier into Afv\'en waves then acoustic waves
(\citeads{1967IAUS...28..429L}); % Lighthill
their amplitude may be a canopy mapper (figure~5 of
\citeads{2010arXiv1012.1196R}).

%%%%%%%%%%%%%%%%%%%%%%%%%%%%%%%%%%%%%%%%%%%%%%%%%%%%%%%%%%%%%%%%%%%%%%%%%%%%
\section{Chromosphere} 
%%%%%%%%%%%%%%%%%%%%%%%%%%%%%%%%%%%%%%%%%%%%%%%%%%%%%%%%%%%%%%%%%%%%%%%%%%%%

%%%%%%%%%%%%%%%%%%%%%%%%%%%%%%%%
\rrsubsection{Overall structure}
%%%%%%%%%%%%%%%%%%%%%%%%%%%%%%%%

\topic{1D modelling}
%%%%%%%%%%%%%%%%%%%%
In the magnetic network shocks arise even deeper.  The fluxtube
simulation by \citetads{1998ApJ...495..468S} % Steiner++ 2D sim
already suggested that photospheric shocks frequently arise in and
near magnetic concentrations, excited by strong adjacent downflows
driven by radiative cooling through the tube walls.  Recently,
\citetads{2011ApJ...730L..24K} % Kato++ fluxtube waves
elaborated the mechanism, calling it magnetic pumping.  The movie
version of figure~\ref{fig:hion2} also shows much shock activity in
and near the magnetic concentrations.  In the simulation of
\citetads{2010ApJ...709.1362L} % Leenaarts++ NaD1
magnetic-concentration shocks occur as low as $h\is 300$\,km and
affect the core of \NaI\ \Done.  The \NaI\ \Done\ observations of
\citetads{2010ApJ...719L.134J} % Jess++ NaD1
and \citetads{2011A&A...531A..17R} % Rutten++ NaD1 revgran
confirm such shock sensitivity.  It makes \NaI\ \Done\ Dopplergrams
effective proxy-magnetograms by marking magnetic concentrations
through the resulting core asymmetry.  This ubiquity of deep-seated
shocks invalidates static 1D modelling even more for network than for
internetwork.

\topic{Chromospheric heating}
%%%%%%%%%%%%%%%%%%%%%%%%%%%%%
Classically, the chromospheric heating budget was determined by
evaluating the net radiative cooling in a standard model (figure~49 of
\citeads{1981ApJS...45..635V}) % VALIII
or determining the amount of excess heating over radiative equilibrium
that is needed to obtain such models
(\citeads{1989ApJ...346.1010A}). % Anderson+AthayII
However,
\citetads{1995ApJ...440L..29C} %C Carlsson+Stein, does chrom T-rise exist?
demonstrated that a cool but shock-ridden clapotisphere delivers the
hot VAL3C chromosphere when fitting ultraviolet continua, due to the
non-linear Planck sensitivity to the hottest phases.  In addition, the
enhanced network brightness in these continua (in which network
appears as in \CaII\ \HK\ filtergrams, bright but without fibrils as
in \Halpha) is a mix of scattered photospheric fluxtube-wall
radiation, brightness from Joule heating
(\citeads{2010MmSAI..81..582C}), % Carlsson++ fluxtubes
and a haze of unresolved high-reaching spicules-II (see below).
Modelling these diverse contributions as radiation loss at the height
of 1D non-fluxtube continuum formation is a severe simplification.
 
The long debate whether acoustic waves provide important
ubiquitous chromosphere heating has `no' as the present answer
(\citeads{2005Natur.435..919F}). % Fossum+Carlsson high-f

%%%%%%%%%%%%%%%%%%%%%%%%%%%%%
\rrsubsection{Fine structure}
%%%%%%%%%%%%%%%%%%%%%%%%%%%%%

\topic{Internetwork fibrils}
%%%%%%%%%%%%%%%%%%%%%%%%%%%%
The long \Halpha\ fibrils that cover (parts of) cell interiors were
called `flow mappers' by Noyes in figure~\ref{fig:Noyes}.  However,
flows along them have been studied only scarcely for quiet-Sun
conditions (\citeads{2009A&A...507.1625C}). % Contarino++ 8542 mottles
Watching \Halpha\ movies gives a strong impression that the fibrils
constitute canopies that ride on the clapotisphere.
\citetads{2008SoPh..251..533R} % Rutten++ tomo7 Monash vanVeelen
found that they partake in shocks coming up underneath, as also
evident in the Doppler timeslices of
\citetads{2009A&A...503..577C}. % Cauzzi++ our IBIS CaHa
The co-aligned image mosaics at
\url{http://www.arcetri.astro.it/science/solare/IBIS/gallery} indicate
that the transition region morphology seen in \HeII\,304\,\AA\ follows
the \Halpha\ fibril pattern.

So far, the MHD simulations do not produce these long fibrils.  Why
not is not clear, perhaps a lack of resolution, extent, radiation
physics, magnetic complexity, and/or history.

The long \Halpha\ fibrils are often taken to outline chromospheric
fields.  Indeed, their filamentary patterns give a vivid impression of
magnetic connectivity.  However, direct correspondence remains
unproven
(\citeads{2011A&A...527L...8D}), % delaCruz+Socas fibril inversion
and the thermal and dynamic fine structure evidenced by the fibrils
may be more complex than the field structure
(\citeads{2006ASPC..354..259J}). % Judge Steinfest
Since \CaII\,854.2\,nm responds to higher temperature with larger
brightness while \Halpha\ does not
(\citeads{2009A&A...503..577C}), % Cauzzi++ our IBIS CaHa,
\Halpha\ brightness may be the better field mapper if \CaII\,854.2\,nm
brightness shows non-aligned features along heating sites.  Also, the
fibril canopies have less opacity in \CaII\,854.2\,nm so that this
line instead shows clapotispheric acoustics in internetwork hearts
(\citeads{2009A&A...494..269V}). % Vecchio++ IBIS shocks

If \Halpha\ fibrils do map chromospheric fields, then extrapolation
using the observed \Halpha\ fibril connectivity as a constraint may help
in nonlinear force-free field prediction of the shape and free-energy
loading of coronal fields from photospheric magnetograms
(\citeads{2008ApJ...672.1209B}; % Bobra+AvanB: Halpha NLFFF
\citeads{2008SoPh..247..249W}). % Wiegelmann++ Halpha NLFFF
If so, space-weather prediction will eventually need continuous
high-resolution full-disk \Halpha\ imaging, best done from space with
large detector mosaics.  Direct polarimetric measurement of
chromospheric fields is an important driver for large solar
telescopes, but reaching quiet-Sun fibril mapping is a daunting
challenge.

\topic{Dynamic fibrils}
%%%%%%%%%%%%%%%%%%%%%%%
Near network and plage shorter fibrils are seen in \Halpha\ as
extending and contracting stalks with abrupt transition-region caps
(\citeads{1995ApJ...450..411S}), % Suematsu++ DFs
and similarly in \Lyalpha\
(\citeads{2009A&A...499..917K}). % Koza++ VAULT
These became a chromospheric success story with the work of
\citetads{2006ApJ...647L..73H} % Hansteen++ DFs
and \citetads{2007ApJ...655..624D}, % DePontieu++ DFs
who identified them as $p$-mode-driven field-guided shocks that slant
up from network and plage.  The slant helps the waves to propagate by
reducing the effective gravity
(\citeads{1973SoPh...30...47M}; % Michalitsanos inclined cutoff
\citeads{1977A&A....55..239B}; % Bel+Leroy inclined cutoff
\citeads{1990LNP...367..211S}). % Suematsu inclined cutoff
They are more upright than the long cell-covering fibrils and seem to
stick out as regular (`type-I') spicules above the dense carpet made
by those at the limb.
 
\topic{Spicules-II\,/\,straws\,/\,RBEs}
%%%%%%%%%%%%%%%%%%%%%%%%%%%%%%%%%%%%%%%%
These features (respectively off-limb\,/\,near-limb\,/\,on-disk) also
border network and plage but they are yet more upright, slenderer,
more dynamic, and reach higher.  They were found in \CaIIH\ near
the limb
%% with the
%% DOT\footnote{Dutch Open Telescope: \url{http://dot.astro.uu.nl}.}  
and called `straws' by me
(\citeads{2006ASPC..354..276R}), % Rutten Steinfest
%BB \citeyearads{2007ASPC..368...27R}), % Rutten Coimbra review
observed as `type-II spicules' in \CaIIH\ off the limb 
%% with Hinode
%%/SOT\footnote{Hinode:
%%  \url{http://hinode.nao.ac.jp/index_e.shtml}; SOT:
%%  \url{http://sot.lmsal.com}.}  
by 
\citeauthor{2007PASJ...59S.655D}
(\citeyearads{2007PASJ...59S.655D}, % BdP++ spicules-II
\citeyearads{2007Sci...318.1574D}), % BdP++ spicules-II = Alfven
and identified on the disk as dark `rapid blue excursions' (RBEs) in the blue
wings of \CaII~854.2~nm and \Halpha\ by
\citetads{2008ApJ...679L.167L} % Langangen++ search disk spicules-II
and \citetads{2009ApJ...705..272R}. % Rouppe++ RBESs
They may also be diagnosed from EUV line
asymmetries
(\citeads{2008ApJ...673L.219M}; % McIntosh++ UV line widths = spicules-II
\citeads{2008ApJ...678L..67H}; % Hara++ upflows
\citeads{2010ApJ...718.1070H}; % Hansteen etal sim downdrafts
\citeads{2011ApJ...732...84M}; % Martinez-Sykora++ EUV asymmetries
\citeads{2011ApJ...738...18T}). % Tian++ EUV asymmetries
They are probably a key player in providing mass and energy to the
corona outside active regions
(\citeads{2009ApJ...701L...1D}). % BdeP++ coronal heating roots
Recently, \citetads{2011Sci...331...55D} % DePontieu++ coronal jets
traced 
%% Hinode/SOT 
RBEs as coronal heating events in
%% SDO/AIA\footnote{Atmospheric Imaging Assembly:
%%  \url{http://aia.lmsal.com}.}  
ultraviolet images.  They appear as jets that send up bullets of hot
gas with precursor fronts at coronal temperatures which continue up
and out while the bullets drop back.  These dynamic features occur
near network and plage all over the Sun, not preferentially in bipolar
areas or coronal holes.  At the limb they appear as long spicules (up
to 10\,Mm, as in figure~\ref{fig:Noyes}) that sway rapidly, likely
from Alfv\'en waves
(\citeads{2007Sci...318.1574D}). % BdP++ spicules-II = Alfven

The recent MHD simulation of a spicule-II-like event by
\citetads{2011ApJ...736....9M} % Martinuez-Sykora++ spicule-II simulation
suggests a tangential field discontinuity in the chromosphere due to
small-scale flux emergence as the principal ingredient, reminiscent of
the squeezed tubes of \citetads{1955ApJ...121..349B}, % Babcocks
{\em ``exciting upward travelling waves [\dots] with tenuous clouds of
  ions and electrons squeezed ahead''\/}.  Perhaps with helical motion
as suggested by
\citetads{2011A&A...531A.173B}. % Beck+Rezai CaIIK limb

\topic{Data inversion}
%%%%%%%%%%%%%%%%%%%%%%
Spline-function inversion techniques loose credibility where
clapotispheric and chromospheric fluctuations are too wild for smooth
1D fitting.  Fibrils are like overlying Schuster-Schwarzschild slabs, and
so the filamentary nature of the chromosphere suggests cloud modelling.
%BB Beckers
%BB Alissandrakis
%BB Heinzel+Schmieder
Inversion techniques using multi-parameter cloud profile synthesis
have been reviewed by
\citetads{2007ASPC..368..217T}; % Tziotziou Coimbra cloud modeling
cloud modelling presently reaches its highest sophistication in
filament-thread analysis (\eg\
%BB Heinzel++
%BB Tsiropoula++
\citeads{2009A&A...503..663G}). % Gouttebroze+Labrosse filament thread
The slowness of hydrogen (and worse for helium)
ionisation/recombination balancing in post-shock gas plus the ubiquity
of frequent shocking requires knowledge of the local history, if not
the wide-area history including the field topography.  Presently,
grasping the physics processes via forward modelling by simulation
with line synthesis has higher priority.

%%%%%%%%%%%%%%%%%%%%%%%%%%%%%%%%%%%%%%%%%%%%%%%%%%%%%%%%%%%%%%%%%%%%%%%%%%%%
\section{Conclusion}
%%%%%%%%%%%%%%%%%%%%%%%%%%%%%%%%%%%%%%%%%%%%%%%%%%%%%%%%%%%%%%%%%%%%%%%%%%%%

The overall and the fine structure of the quiet photosphere are largely
understood.  The structure of the clapotisphere is reasonably
understood.  The fine structure of the quiet chromosphere (of which
the spatiotemporal mean is not of interest) is not understood
excepting dynamic fibrils, but holds the key to the mass and energy
loading of the quiet-Sun outer atmosphere.  The upcoming large groundbased
telescopes, ultraviolet space missions and increased realism of MHD
simulations should combine into substantial progress in this premier
solar physics frontier.

Because open fields (in a local sense) harbour more upward
connectivity than closed fields, they are the ones to concentrate on
even though the corresponding fine structure is harder to observe.  In
the quiet-Sun photosphere these are the kilogauss field
concentrations, in the quiet-Sun chromosphere
spicules-II\,/\,straws\,/\,RBEs.  \Halpha\ mapping of closed fields
(within the chromosphere) appears especially useful for eruptivity
prediction.

%%%%%%%%%%%%%%%%%%%%%%%%%%%%%%%%%%%%%%%%%%%%%%%%%%%%%%%%%%%%%%%%%%%%%%%%%%%%
\begin{acknowledgements}
  I thank J.~Leenaarts and N.~Vitas for assistance, the referees for
  making me rewrite an earlier draft, and the issue editor for
  allowing additional reference pages.
\end{acknowledgements}

%%%%%%%%%%%%%%%%%%%%%%%%%%%%%%%%%%%%%%%%%%%%%%%%%%%%%%%%%%%%%%%%%%%%%%%%%%%%
%% references
%RE the highly substandard rsguide does not even mention BibTeX...
%RE rspublicnat.bst from http://links.tedpavlic.com/bst/rspublicnat.bst
%RR what a lousy space-eating way of setting references most superfluously

%%\bibliographystyle{rspublicnat}

%RE the below settings are necessary because the "thebibliography" definition 
%RE of rspublic.cls is overwritten by natbib.sty 
%%\begin{small} \bibsep=0pt
 %% \bibliography{journals,rjrfiles,adsfiles} %RR jrj first for ADS repairs

%RR used AA style instead of the silly Royal Society style
%RR edited: ASPCS, Am Inst physics, Vol out, endpages out, Lites thesis

\bibliographystyle{aa}
\begin{footnotesize} \bibsep=0pt

%%  \bibliography{aajour,rjrfiles,adsfiles} %RR rjr first for ADS corrections

\end{footnotesize}

\end{document}